\documentclass[12pt]{article}
\begin{document}
%\preprint{UTPT-99-02}
\begin{center}
{\large\bf M-Theory}
\vskip 0.3 true in
{\large J. W. Moffat}
\vskip 0.3 true in
{\it Department of Physics, University of Toronto,
Toronto, Ontario M5S 1A7, Canada}
\end{center}
%\date{\today}
\begin{abstract}%
We construct an eleven-dimensional superspace with superspace coordinates
and formulate a finite M-theory using non-anticommutative geometry. The
conjectured M-theory has the correct eleven-dimensional supergravity low
energy limit. We consider the problem of finding a stable finite M-theory
which has de Sitter space as a natural ground state, and the problem of
eliminating possible future horizons.
\end{abstract}
\vskip 0.2 true in
e-mail: moffat@medb.physics.utoronto.ca

%\pacs{ }

\section{Introduction}

String duality theory has brought about the return of eleven-dimensional
supergravity. The strong-coupling limit of the low-energy sector of type
IIA superstring theory is eleven-dimensional supergravity.
Eleven-dimensional supergravity is twenty three years old~\cite{Julia} and
by itself does not constitute a consistent unification of quantum gravity
and supersymmetry, because the ubiquitous divergences of weak field quantum
gravity persist. On the other hand, superstring theory is claimed to
consistently unify gravity and quantum mechanics into a finite theory of
quantum gravity. The question then arises: What is the consistent strong
coupling theory in eleven dimensions that contains eleven-dimensional
supergravity as its low energy limit? Since M-theory does not itself
contain strings, perhaps it and its low-energy limit may not have
anything directly to do with string theory or D-brane theory. Since we do
not know the degrees of freedom of M-theory, we must demand certain
consistency requirements of M-theory:

\begin{enumerate}
\item It must support a finite quantum gravity theory,

\item The theory must reduce to either a massless or
massive form of eleven-dimensional supergravity in a low energy limit,

\item The fermions must have a chiral low energy limit,

\item It should contain the standard model,

\item It should describe a realistic cosmology in curved
spacetime.
\end{enumerate}

After almost thirty years of study in supersymmetry theory, there is still
no experimental evidence that it plays a role in nature. It is the {\it
uniqueness} of eleven-dimensional supergravity that makes it such an
attractive proposal for a unification of gravity and quantum mechanics and
as a low energy limit of an M-theory. Standard Kaluza-Klein theories do not
possess a similar uniqueness, since a large number of dimensions and group
structures can potentially qualify as the constituents of a unified theory.
Moreover, the indirect evidence that the coupling constants of the minimal
supersymmetric standard model (MSSM) meet at a unifying energy $\sim
10^{16}$ GeV, does give some credence to the validity of the MSSM theory.

Little is known about the nature of M-theory, but it has allowed us to
extend our knowledge of string theory beyond its applicability. Banks,
Fischler, Shenker and Susskind~\cite{Banks} conjectured that the
microscopic degrees of freedom of M-theory, when pictured in an infinite
momentum Lorentz frame, are $D_0$-branes. They described the dynamics of
the eleven-dimensional space by a $N\times N$ matrix quantum mechanics.
They conjectured that M-theory is equivalent to a matrix quantum mechanics
of $U(N)$ matrices in the $N\rightarrow\infty$ limit with a Hamiltonian
that follows from reducing $9+1$-dimensional $U(N)$ super Yang-Mills theory
to $0+1$ dimensions. Horava and Witten~\cite{Horava} developed a heterotic
M-theory by compactifying eleven-dimensional theory on an $S^1/Z_2$
orbifold corresponding to the strong coupling limit of heterotic
ten-dimensional $E_8\times E_8$ string theory. Compactifying an additional
six dimensions on a Calabi-Yau 3-manifold leads, in the low energy limit,
to four-dimensional $N=1$ supersymmetry theory.

In the following, we shall study
possible models of M-theory which do not owe their finiteness to string
theory but to an eleven-dimensional superspace that involves
noncommutative as well as non-anticommutative coordinates with a
${\hat f}\circ {\hat g}$ product of operators ${\hat f}$ and
${\hat g}$ on a Hilbert superspace~\cite{Moffat,Moffat2,Moffat3}. The
problem of the physical necessity for chiral fermion fields is solved by an
orbifold compactification with periodic boundary conditions on the fermion
field operators, or, alternatively, by compactifying the finite M-theory
along a noncompact direction.

Any realistic M-theory must be consistent with modern cosmological data.
This means that it must describe a curved universe with positive but small
cosmological constant and be consistent with the now overwhelming data
supporting an accelerating universe~\cite{Perlmutter}. This immediately
poses a problem for M-theory (and string theory), because the standard
M-theory cannot contain a positive cosmological constant. De Sitter space
cannot be contained in standard eleven-dimensional supergravity or
M-theory. The de Sitter superalgebra is not contained in M-theory and in a
supersymmetric theory the de Sitter space solutions have zero energy and,
indeed, there is no positive energy theorem. In view of these difficulties,
we shall also consider a recent variant of M-theory call MM-theory (massive
M-theory)~\cite{Lambert}.

Recently, another difficulty in formulating a consistent string theory or
M-theory has arisen with the possibility of a future horizon, associated
with an eternally accelerating universe based either on a positive
cosmological constant or a quintessence model. The future horizon forbids
the construction of a consistent S-matrix formalism based on asymptotic in
and out states at infinity~\cite{Susskind,Witten}.
Various possible solutions to this problem have been
proposed~\cite{Moffat4,Cline}.

\section{\bf Eleven-Dimensional Superspace}

Let us define an eleven-dimensional superspace with the superspace
coordinates~\cite{Moffat}:
\begin{equation}
\rho^M=x^M+\beta^M,
\end{equation}
where $M=0,1,...10$ and the $x^M$ denote the classical commuting
c-number coordinates of the space, $[x^M,x^N]=0$, and $\beta^M$ denote
the anticommuting Grassmann coordinates, $\{\beta^M,\beta^N\}=0$.

Both noncommutative and non-anticommutative geometries can be unified
within the superspace formalism using the $\circ$-product of two
operators ${\hat f}$ and ${\hat g}$:
\begin{equation}
\label{fgproduct}
({\hat f}\circ{\hat g})(\rho)
=\biggl[\exp\biggl(\frac{1}{2}\omega^{MN}
\frac{\partial}{\partial\rho^M}\frac{\partial}{\partial\eta^N}\biggr)
f(\rho)g(\eta)\biggr]_{\rho=\eta} $$ $$
=f(\rho)g(\rho)+\frac{1}{2}\omega^{MN}\partial_M
f(\rho)\partial_Ng(\rho)+O(\omega^2),
\end{equation}
where $\partial_M=\partial/\partial\rho^M$ and $\omega^{MN}$ is a
nonsymmetric tensor
\begin{equation}
\omega^{MN}=-\tau^{MN}+i\theta^{MN},
\end{equation} with
$\tau^{MN}=\tau^{NM}$ and $\theta^{MN}=-\theta^{NM}$.
Moreover, $\omega^{MN}$ is Hermitian symmetric
$\omega^{MN}=\omega^{\dagger MN}$, where $\dagger$ denotes Hermitian
conjugation. The familiar commutative coordinates of spacetime are replaced
by the superspace operator relations
\begin{equation}
\label{commutator}
[{\hat\rho}^M,{\hat\rho}^N]=2\beta^M\beta^N+i\theta^{MN},
\end{equation}
\begin{equation}
\label{anticommutator}
\{{\hat\rho}^M,{\hat\rho}^N\}=2x^Mx^N+2(x^M\beta^N+x^N\beta^M)-\tau^{MN}.
\end{equation}
In the limit that $\beta^M\rightarrow 0$ and
$\vert\tau^{MN}\vert\rightarrow 0$, we get the familiar expression for
noncommutative coordinate operators
\begin{equation}
[{\hat x}^M,{\hat x}^N]=i\theta^{MN}.
\end{equation}
In the limits $x^M\rightarrow 0$ and $\vert\theta^{MN}\vert\rightarrow 0$,
we obtain the Clifford algebra anticommutation relation
\begin{equation}
\{{\hat\beta}^M,{\hat\beta}^N\}=-\tau^{MN}.
\end{equation}

In the following, we shall consider the simpler non-anticommutative
geometry obtained in the limit $\theta^{MN}=0$, because it alone can lead
to a finite and unitary quantum field theory and quantum gravity theory
~\cite{Moffat,Moffat2,Moffat3}. In the non-anticommutative field theory
formalism, the product of two operators ${\hat f}$ and ${\hat g}$ has a
corresponding $\diamondsuit$-product
\begin{equation}
({\hat f}\diamondsuit
{\hat g})(\rho)
=\exp\biggr(-\frac{1}{2}\tau^{MN}\frac{\partial}{\partial\rho^M}\frac{\partial}
{\partial\eta^N}\biggr)f(\rho)g(\eta)\vert_{\rho=\eta}
$$ $$
=f(\rho)g(\rho)-\frac{1}{2}\tau^{MN}\partial_M f(\rho)\partial_N
g(\rho)+O(\tau^2).
\end{equation}

\section{\bf M-theory}

In previous work~\cite{Moffat5}, it was shown that
noncommutative quantum field theory cannot give a renormalizable
quantum gravity. On the other hand, the non-anticommutative superspace
quantum field theory can lead to a finite perturbative quantum gravity
theory~\cite{Moffat,Moffat2,Moffat3}. We shall apply this quantum field
theory formalism to M-theory. We begin with the standard eleven-dimensional
supergravity, describing the highest number of dimensions in which
supersymmetry representations with $J\leq 2$ can exist. Its reduction to
four dimensions is automatically guaranteed to give an $O(8)$ invariant
supergravity theory with $N=8$ supersymmetry.

The field content consists of the vielbein $e^A_M$ (where $A,B,C...$ refer
to tangent space indices), a
Majorana spin $\frac{3}{2}$ $\psi_M$, and of a completely antisymmetric
gauge tensor field $A_{MNP}$. The metric is $(+---...-)$ and the
eleven-dimensional Dirac matrices satisfy
\begin{equation}
\{\Gamma_A,\Gamma_B\}=-2\eta_{AB},
\end{equation}
where $\eta_{AB}$ denotes the flat Minkowski tangent space metric.
Moreover, $\Gamma^{A_1...A_N}$ denotes the product of $N\Gamma$ matrices
completely antisymmetrized.

Our superspace M-theory Lagrangian,
using the $\diamondsuit$-product has the form
\begin{equation}
\label{Lagrangian} {\cal
L}=-\frac{1}{4\kappa^2}e\diamondsuit R(\omega)
-\frac{i}{2}e\diamondsuit
{\bar\psi}_M\diamondsuit\Gamma^{MNP}D_N\biggl(\frac{\omega+{\hat\omega}}{2}\biggr)
\diamondsuit\psi_P-\frac{1}{48}e\diamondsuit F_{MNPQ}\diamondsuit F^{MNPQ}
$$ $$
+\frac{\kappa}{192}
e\diamondsuit
({\bar\psi}_M\diamondsuit\Gamma^{MNOPQR}\psi_N+12{\bar\psi}^P\diamondsuit
\Gamma^{OR}\psi^Q)\diamondsuit(F_{PQOR}+{\hat F}_{PQOR})
$$ $$
+\frac{2\kappa}{(144)^2}\epsilon^{O_1O_2O_3O_4P_1P_2P_3P_4MNR}F_{O_1O_2O_3O_4}
\diamondsuit F_{P_1P_2P_3P_4}\diamondsuit A_{MNR}, \end{equation} where
$R(\omega)$ is the scalar contraction of the curvature tensor
\begin{equation}
R_{MNAB}=\partial_M\omega_{NAB}-\partial_N\omega_{MAB}+{\omega_{MA}}^C\diamondsuit\omega_{NCB}
-{\omega_{NA}}^C\diamondsuit\omega_{MCB},
\end{equation}
and $F_{MNOP}$
is the field strength defined by
\begin{equation}
F_{MNOP}=4\partial_{[M}A_{NOP]},
\end{equation}
with $[...]$ denoting the
antisymmetrized sum over all permutations, divided by their number.

The covariant derivative is
\begin{equation}
D_N(\omega)\psi_M=\partial_N\psi_M+\frac{1}{4}\omega_{NAB}\diamondsuit\Gamma^{AB}\psi_M.
\end{equation}
The spin connection $\omega_{MAB}$ is defined by
\begin{equation}
\omega_{MAB}=\omega^0_{MAB}(e)+T_{MAB},
\end{equation}
where $T_{MAB}$ is the spin torsion tensor.

The transformation laws are
\begin{equation}
\delta e^A_M=-i\kappa{\bar\epsilon}\diamondsuit\Gamma^A\psi_M,
\end{equation}
\begin{equation}
\delta\psi_M=\frac{1}{\kappa}D_M({\hat\omega})\diamondsuit\epsilon+\frac{i}{144}
({\Gamma^{OPQR}}_M -8\Gamma^{PQR}\delta^O_M)\epsilon\diamondsuit{\hat
F}_{OPQR} \equiv \frac{1}{\kappa}{\hat D}_M\epsilon,
\end{equation}
\begin{equation}
\delta A_{MNP}=\frac{3}{2}{\bar\epsilon}\diamondsuit\Gamma_{[MN}\psi_{P]},
\end{equation}
where
\begin{equation}
{\hat F}_{MNPQ}=F_{MNPQ}
-3\kappa{\bar\psi}_{[M}\diamondsuit\Gamma_{NP}\psi_{Q]}.
\end{equation}
We also have
\begin{equation}
{\hat\omega}_{MAB}=\omega_{MAB}+\frac{i\kappa^2}{4}{\bar\psi}_O
\diamondsuit{\Gamma_{MAB}}^{OP}\psi_P.
\end{equation}

In the limit $\vert\tau^{MN}\vert\rightarrow 0$ and
$\beta^M\rightarrow 0$, (\ref{Lagrangian}) reduces to the Cremmer, Julia
and Scherk (CJS) eleven-dimensional supergravity Lagrangian~\cite{Julia},
which should be the correct low energy limit of an M-theory. The finiteness
and gauge invariance of the M-theory should be guaranteed by the
non-anticommutative field theory. However, this finiteness was proved for
scalar field theory and weak field quantum
gravity~\cite{Moffat,Moffat2,Moffat3}, but is expected to hold also for our
non-perturbative M-theory, due to the existence of a finite, fundamental
length $\ell$. The symmetric tensor $\tau^{MN}$ can be written as
\begin{equation} \tau^{MN}=\ell^2s^{MN}=\frac{1}{\Lambda^2}s^{MN},
\end{equation} where $\Lambda$ is a fundamental energy scale that can be
chosen to be the Planck energy $\Lambda=M_{PL}$.

The focus of experimental cosmology has been on the now significant data
that indicates that the universe is undergoing an accelerated
expansion~\cite{Perlmutter}. It has been know for a long
time~\cite{Duff} that spontaneous compactification of the
eleven-dimensional supergravity field equations leads to solutions with a
vacuum state corresponding to the product of a four-dimensional anti-de
Sitter space with negative cosmological constant and a seven-dimensional
Einstein space with positive cosmological constant. Unfortunately, this
direct compactification leads us to an unacceptable four-dimensional
cosmology. There exist no-go theorems which forbid attempts to realize de
Sitter space within string theory and M-theory~\cite{Maldacena}.

Chamblin and Lambert (CL) have extended standard
eleven-dimensional supergravity theory to a massive supergravity theory, in
which de Sitter space is a natural ground state~\cite{Lambert}. The CL
scenario may not have anything directly to do with standard
string theory, but within our construction of M-theory this is not
important. The important issue is that it may provide a natural embedding
of de Sitter space into our modified eleven-dimensional supergravity or
M-theory.

The CL modification of supergravity theory is based on a conformal spin
$CSpin(1,10)$ connection, defined in our superspace by
\begin{equation}
\Upsilon_A=\frac{1}{4}{\Omega_A}^{BC}\Gamma_{BC}+2k_A,
\end{equation}
where
\begin{equation}
{\Omega_{AB}}^C={\omega_{AB}}^{C}+2(e^C_A\diamondsuit
k_B-e_{AB}\diamondsuit k^C),
\end{equation}
and for which the conformal part of
the curvature $dk$ vanishes. If $k=d\chi$, then the redefinition that
returns the equations of motion back to their usual ones is
\begin{equation}
e^A_M\rightarrow\exp(-2\chi)\diamondsuit e^A_M,\quad
\psi_M\rightarrow \exp(-2\chi)\diamondsuit\psi_M.
\end{equation} This
modification is non-trivial for a non-simply connected space.

Choosing to compactify on the non-simply connected supermanifold,
$SM_{10}\times S^1$, results in the non-anticommutative ten-dimensional
supergravity equations of motion~\cite{Lambert}
\begin{equation}
R_{ab}-\frac{1}{2}g_{ab}\diamondsuit R
=-2[D_a\diamondsuit D_b\phi-g_{ab}\diamondsuit D_c\diamondsuit D^c\phi
+g_{ab}\diamondsuit D_c\phi\diamondsuit D^c\phi] $$ $$
+\frac{1}{2}(F_{ac}\diamondsuit {F_b}^c-\frac{1}{4}g_{ab}\diamondsuit
F^{cd}\diamondsuit F_{cd})\diamondsuit\exp(2\phi) $$ $$
-18m(D_{(a}A_{b)}-g_{ab}\diamondsuit D^cA_c)-36m^2(A_a\diamondsuit
A_b+4g_{ab}\diamondsuit A^c\diamondsuit A_c) $$ $$
-12mA_{(a}\diamondsuit\partial_{b)}\phi-30mg_{ab}\diamondsuit
A^c\diamondsuit\partial_c\phi-144m^2g_{ab}\diamondsuit\exp(-2\phi),
\end{equation} $$ $$
\begin{equation} D^bF_{ab}=18mA_b\diamondsuit
{F_a}^b+72m^2\exp(-2\phi)\diamondsuit
A_a-24m\exp(-2\phi)\diamondsuit\partial_a\phi, \end{equation}
\begin{equation}
6D^a\diamond D_a\phi-8D_a\phi
\diamondsuit D^a\phi=-R+\frac{3}{4}\exp(2\phi)\diamondsuit
F^{ab}\diamondsuit F_{ab} $$
$$ +360m^2\exp(-2\phi)+288m^2A^a\diamondsuit A_a
+96mA^b\diamondsuit\partial_b\phi-36mD^bA_b,
\end{equation}
where $a,b=0,1,...9$ and $F_{ab}=\partial_aA_b-\partial_bA_a$.

These equations of motion were obtained by turning off the four-form field
strength and fermions.
They correspond to the eleven-dimensional equations of motion with no
dependence on the y coordinate with $k=mdy$, where $dy$ is the tangent
vector to the circle. Thus, solutions of these equations of motion are
solutions to our finite M-theory. In the limits that
$\vert\tau^{ab}\vert\rightarrow 0$ ($\ell\rightarrow 0$) and
$m\rightarrow 0$, we recover the standard massless type IIA ten-dimensional
supergravity and the relation of M-theory to perturbative string theory. A
dimensional reduction of eleven-dimensional supergravity with our
$\diamondsuit$-product modification over a {\it noncompact dimension},
yields the same equations of motion~\cite{Pope}.

The compactification of the eleven-dimensional M-theory on a circle has
built into it a mass generating mechanism, such that the two-form eats the
scalar, the three-form eats the vector and the four-form eats the
three-form. However, the $U(1)$ symmetry of the vector $A_a$ is violated,
so that the vector is tachyonic and the solutions will be unstable. Even
though the equations of motion are supersymmetric, there is no Noether
theorem conserved supercharge associated with the Hamiltonian of the
massive theory. Thus, massive M-theory compactified on $S^1$ with a
topologically non-trivial conformal connection does not possess a globally
conserved supercharge.

The massive M-theory does not in its present form possess an action. The
equations of motion come from gauging the scale symmetry of
modified eleven-dimensional supergravity, and whereas the equations of
motion of CL theory obey this gauge symmetry, the action does not.
This is an unusual situation, since normally after a spontaneous symmetry
breaking the modified action exists, given a suitable spontaneous symmetry
breaking potential. Clearly this issue requires further study.

The fact that the massive M-theory is embedded in a de Sitter
space can be seen by turning off all the gauge potentials to give
\begin{equation}
R_{ab}=36m^2\exp(-2\phi)\diamondsuit g_{ab},
\end{equation}
which corresponds to a constant dilaton field $\phi$. In the
limit $\ell\rightarrow 0$ we recover ten-dimensional de Sitter space with
an effective cosmological constant
\begin{equation}
\lambda=576m^2\exp(-2\phi).
\end{equation}
As pointed out
in~\cite{Lambert}, there is no need to consider other fields to induce a
cosmological constant, for a positive cosmological constant asserts itself
in ten dimensions. In de Sitter space we have $\sum\{Q,Q^*\}=0$ and there is
no positive energy theorem.

The question remains to be asked: Do there exist any stable solutions to
the massive M-theory with a natural embedding in de Sitter space?

The problem of obtaining fermions possessing chirality can be resolved in
two ways: (a) by performing a compactification on a {\it noncompact}
direction, as shown by Wetterich~\cite{Wetterich}, or, (b) by performing an
orbifold compactification with the resulting periodic boundary conditions
on the fermion fields leading to chirality of the fermions~\cite{Georgi}.
We have aleady observed that the massive eleven-dimensional supergravity
theories give massive type IIA ten-dimensional supergravity theories when
they are compactified along noncompact directions~\cite{Pope}. The Witten
chirality index theorem for Kaluza-Klein theories only holds for {\it
compact} Riemannian manifolds~\cite{Witten2}. Our finite M-theory can
reduce to a standard model with chiral fermions in four dimensions when
either of (a) or (b) compactifications are performed. The
eleven-dimensional manifold of M-theory can contain the $SU(3)\times
SU(2)\times U(1)$ standard model~\cite{Witten3}.

The finite M-theory we have constructed is a {\it nonlocal} quantum field
theory in eleven-dimensions. The nonlocal nature of the theory will
persist after compactification for $\Lambda < \infty$ ($\ell \not= 0$), and
will guarantee that the quantum gravity sector of the four-dimensional
theory is finite to all orders of perturbation theory~\cite{Moffat3}. That
M-theory is nonlocal should not come as a surprise, for it is now generally
accepted that string and D-brane theories are intrinsically nonlocal
theories. This is the case for noncommutative string and
D-brane theories on a $B$-field background. The nonlocality of the fields
will only occur at short distances and will vanish as
$\Lambda\rightarrow\infty$.

If we have succeeded in finding a finite M-theory with our
non-anticommutative geometry and can guarantee that it is naturally
embedded in de Sitter space with stable solutions, then we are faced with
the problem of a future horizon, either through a positive cosmological
constant or through a quintessence-like equation of
state~\cite{Susskind}. For a positive cosmological constant, the
universe will undergo eternal acceleration, whereas for a quintessence dark
energy, it may be possible to construct a quintessence model of cosmology
in which the universe will begin to decelerate in the future, thereby
avoiding the existence of a future horizon~\cite{Cline}. The existence of a
future horizon would make our formulation of finite M-theory incompatible
with the existence of an S-matrix and consistent physical observables.
Particles would be immersed in a heat bath with a finite entropy and there
would not exist asymptotic in and out states at infinity. The number of
degrees of freedom would be finite and the dimensions of our Hilbert
superspace would also be finite, which would invalidate the
basic notions of our non-anticommutative and noncommutative quantum field
theory.

One way to resolve the problem of future horizons is to postulate that the
speed of light varies in the future as well as in the past, for with a
suitable varying speed of light as the universe expands,
any future horizon can be eliminated, allowing for an infinite spacetime
with appropriate in and out states at infinity and a consistent
S-matrix~\cite{Moffat4}. A varying speed of light is a natural outcome of
higher-dimensional theories as is the existence of multiple light cones,
which can undergo expansion or contraction and remove future
horizons~\cite{Clayton,Drummond,Liberati}.

\section{Conclusions}

We have formulated an
eleven-dimensional superspace with the algebra of functions on
a noncommutative and non-anticommutative space
isomorphic to the algebra of functions with commutative $x^M$ and
anticommutative $\beta^M$ coordinates, with the general ${\hat
f}\circ{\hat g}$-product of operators ${\hat f}$ and ${\hat g}$. We
constructed a conjectured finite M-theory, using the simpler
non-anticommutative geometry with the ${\hat f}\diamondsuit{\hat
g}$-product and eleven-dimensional
supergravity theory. This M-theory should produce a finite quantum gravity
theory and Yang-Mills gauge theory coupled to the Majorana spin
$\frac{3}{2}$ fermion field. In the limit $\vert\tau^{MN}\vert\rightarrow
0$, this finite M-theory reduces to eleven-dimensional CJS supergravity
theory~\cite{Julia}, which is the correct low energy effective theory of
M-theory, related by duality to type IIA superstring theory.

Demanding that the finite M-theory be naturally embedded in a de Sitter
space with positive cosmological constant $\lambda >0$, led us to
formulate a non-anticommutative geometrical, massive M-theory based on the
CL~\cite{Lambert} massive supergravity theory.
The field equations of this theory reduce in the limits
$\beta^M\rightarrow 0$, $\vert\tau^{MN}\vert\rightarrow 0$ and
$m\rightarrow 0$ to massless type IIA supergravity theory, formulated in
flat Minkowski space. However, this theory may not possess stable
solutions, although it is possible that it could relax to a true ground
state which is stable and is still embedded in de Sitter space. These
issues require further investigation.

The problem of the existence of a future horizon in a de Sitter space
can be removed by postulating that the speed of light varies in the future
universe as well as in the past universe~\cite{Moffat4}. Multiple expanding
or contracting light cones, existing in the higher-dimensional theory as
well as in the compactified theory, can remove a future horizon and allow
for a consistent S-matrix formulation of our finite
M-theory~\cite{Clayton,Drummond,Liberati}.

\vskip 0.2 true in
{\bf Acknowledgments}
\vskip 0.2 true in
I thank Neil Lambert for helpful and
stimulating conversations and correspondence. This work was supported by
the Natural Sciences and Engineering Research Council of Canada.
\vskip 0.5 true in

\end{document}